\documentclass[prl,twocolumn,showpacs,floatfix]{revtex4}

\usepackage{graphicx}
\begin{document}

\title{Enhancement of the superconducting transition temperature in Nb/Permalloy
bilayers by controlling the domain state of the ferromagnet.}

\author{A. Yu. Rusanov}\affiliation{Kamerlingh Onnes Laboratory,
  Universiteit Leiden, PO Box 9504, 2300 RA Leiden, The Netherlands}

\author{M. Hesselberth}\affiliation{Kamerlingh Onnes Laboratory,
  Universiteit Leiden, PO Box 9504, 2300 RA Leiden, The Netherlands}

\author{J. Aarts}\affiliation{Kamerlingh Onnes Laboratory,
  Universiteit Leiden, PO Box 9504, 2300 RA Leiden, The Netherlands}

\author{A. I. Buzdin}\affiliation{CPMOH, U. M. R. 5798, Universit\'{e} Bordeaux I, 33405 Talence
Cedex, France}

\date{\today}

\begin{abstract}
In (S/F) hybrids the suppression of superconductivity by the exchange field $h_{ex}$ of
the ferromagnet can be partially lifted when different directions of $h_{ex}$ are sampled
simultaneously by the Cooper pair. In F/S/F trilayer geometries where the magnetization
directions of the two F-layers can be controlled separately, this leads to the so-called
spin switch. Here we show that domain walls in a single F-layer yield a similar effect.
We study the transport properties of Ni$_{0.80}$Fe$_{0.20}$/Nb bilayers structured in
strips of different sizes. For large samples a clear enhancement of superconductivity
takes place in the resistive transition, in the very narrow field range (order of 0.5~mT)
where the magnetization of the Py layer switches and many domains are present. This
effect is absent in microstructured samples. Comparison of domain wall width $\delta_w$
to the temperature dependent superconductor coherence length $\xi_S(T)$ shows that
$\delta_w \approx \xi_S(T)$, which means that the Cooper pairs sample a large range of
different magnetization directions.
\end{abstract}

\pacs{74.45.+c,74.78.-w,85.25.Hv}

\maketitle

\begin{figure}
{\includegraphics[width=7cm]{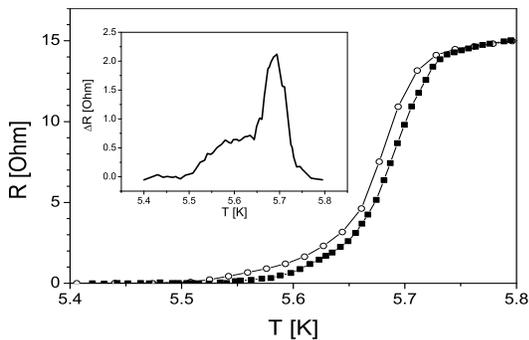}} \caption{Resistance $R$ versus
temperature $T$ of the 'large' sample s/Py(20)/Nb(21) in zero field (open symbols) and in
a field of 4.2~mT (filled symbols). The inset shows the difference between the two data
sets. } \label{f1-RT}
\end{figure}

\begin{figure}
{\includegraphics[width=7.5cm]{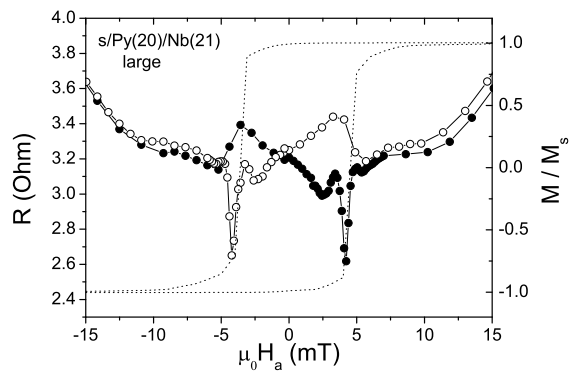}} \caption{Left-hand scale :
Resistance $R$ versus applied field $\mu_0 H_a$ of the 'large' sample s/Py(20)/Nb(21).
Filled symbols are in positive field (forward) direction, open symbols in backward
direction. Right-hand scale : magnetization $M$ (dotted lines) normalized on the
saturation magnetization $M_{s}$ versus $\mu_0 H_a$, measured at 8~K. In both cases, $H_a
\|$ \^e$_{e}$ (the easy axis of magnetization).} \label{f2-RHRM}
\end{figure}

\begin{figure}[b]
{\includegraphics[width=8cm]{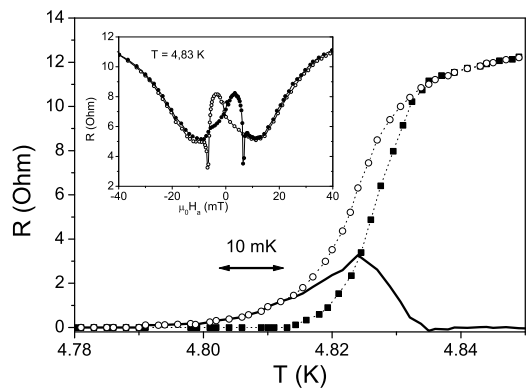}} \caption{Resistance $R$ vs.
temperature $T$ for a 'large' sample s/Nb(19)/Py(20) in zero field (open symbols) and in
a field of 6.3~mT (filled symbols). Dotted lines are guides to the eye, the drawn line is
the difference between the two data sets. Inset : $R$ as function of applied field $\mu_0
H_a$. Filled symbols are in forward direction, open symbols in backward direction.}
\label{f3-P20N20}
\end{figure}

\begin{figure}[t]
{\includegraphics[width=8cm]{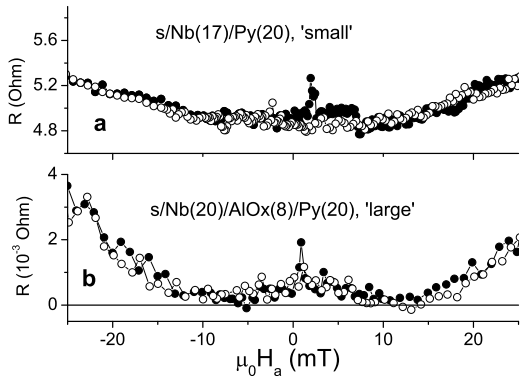}} \caption{Resistance $R$ vs. magnetic
field $\mu_0H_a$ (a) for a 'small' sample s/Nb(17)/Py(20); (b) same for a large sample
s/Nb(20)/AlOx(8)/Py(20). Filled symbols are in forward direction, open symbols in
backward direction.} \label{f4-alox}
\end{figure}

\begin{figure}
{\includegraphics[width=6cm]{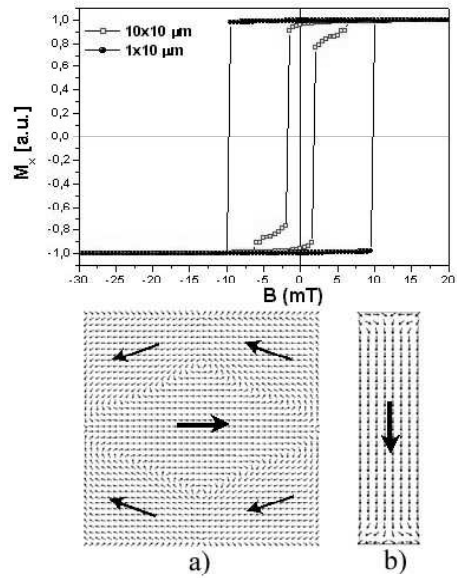}} \caption{The switching
behavior of small Py structures. The top picture represents magnetization vs. external
magnetic field for 10$\mu$m long Py structures with aspect ratio \textit{a}=1 (open
squares, narrow loop) and \textit{a}=10 (closed squares, wide loop). The bottom picture
shows the correspondent domain configuration near the coercive field for both structures.
For samples with the aspect ratio above 2.5 we found no stable multi-domain state during
magnetization switching.} \label{fig5-domains}
\end{figure}

Proximity effects between a superconductor (S) and a ferromagnet (F) are the focus of
much current research, basically because of the possibilities for several distinct and
unusual phenomena. One is due to the fact the exchange field $h_{ex}$ in the ferromagnet
gives rise to an oscillatory damped amplitude of the pairing function. In an S/F/S
geometry, this oscillation allows coupling of the superconducting banks with a phase
change of $\pi$ rather than 0 \cite{rado91,dem97}. This can be witnessed in a
nonmonotonic variation of the superconducting transition temperature $T_c$ as function of
the F-layer thickness $d_F$ \cite{tagir02,laz00}; or, especially with a weak ferromagnet,
in the Josephson current of an S/F/S junction \cite{ryaz01,kontos02} or a superconducting
quantum interference device \cite{guich03}. Other phenomena are linked to a situation in
which the Cooper pair can sample different directions of $h_{ex}$ within its coherent
volume. The best known example is an F/S/F geometry with a thin S-layer, in which the
magnetization of one F-layer can be rotated with respect to the other. The suppression of
the order parameter in S will then be larger when both magnetizations are parallel (P)
and smaller when they are antiparallel (AP). This so-called spin switch has been
discussed theoretically \cite{tagir99,buzdin99} and a first experimental realization was
recently reported by Gu {\it et al.} \cite{gu02}. They used a weak ferromagnet (CuNi) and
reported a small but measurable increase of $T_c^{AP}$ with respect to $T_c^P$ (the
transition temperatures of the P and AP configurations, respectively) of about 5~mK.
However, the physics of the problem is more general. A weak ferromagnet is not an a
priori condition for the effect, and it might even be argued (although this has not been
emphasized in theoretical treatments) that ferromagnets with large exchange fields are
preferable. Also, the trilayer configuration is not the only one which can invoke
different directions for $h_{ex}$ : any domain wall in the ferromagnet offers different
directions intrinsically, both in the wall and on either side. In claiming a coupling
effect between two F-layers, it might even be necessary to check the absence of in-plane
domain wall effects. Moreover, other mechanisms may be at play; inhomogeneous exchange
fields are predicted to induce enhanced superconductivity by spin-triplet excitations
\cite{kadi01,berge01}. \\
In this Letter, we investigate S/F bilayers and show that the domain state of a magnet
with strong spin polarization (Ni$_{80}$Fe$_{20}$, Permalloy (Py)) gives rise to an even
slightly larger increase in $T_c$ of the superconductor, in our case Nb, than mentioned
above. We give a qualitative discussion of the possible mechanism for the effect, which
is basically due to a lowered average exchange field seen by the Cooper pair and might
either be called an in-plane spin switch or domain wall induced $T_c$-enhancement. \\
Samples of Nb/Py were prepared by sputter deposition in an ultrahigh vacuum system. They
were structured in simple bars of either 0.5~mm $\times$ 4~mm ('large' sample) or
1.5~$\mu$m $\times$ 20~$\mu$m ('small' sample). Contacts were not included in the
geometry in order to minimize problems with stray fields from contact pads or arms.
Instead, Au contacts for measuring in 4-point geometry were added by sputter deposition.
The choice of Py as ferromagnet is dictated by the wish for large $h_{ex}$ (the spin
polarization is 45~\% \cite{Mood98}), but equally by the need of well-defined
magnetization switching at low fields. Since an easy axis for magnetization is induced by
the residual magnetic fields in the sputtering machine, care was taken to align the long
axis of the bars with the easy axis of magnetization \^e$_{e}$. Magnetic fields were
applied in the plane of the sample, along the bars and therefore along \^e$_{e}$.
Different layer thicknesses were used for both Nb and Py, which yielded similar results.
We will concentrate on samples with both the Nb and the Py thickness around 20 nm.\\
The zero-field resistance of a 'large' sample of 21 nm Nb on 20 nm Py on a Si substrate
(denoted s/Py(20)/Nb(21)) as function of temperature is shown in Fig.~\ref{f1-RT}. The
transition temperature is around 5.7~K, depressed from the pure Nb value by the proximity
of the F-layer. The width is about 100~mK.
After stabilizing the temperature at around 5.65~K, in the transition region, the
magnetic field $\mu_0H_a$ is swept from 65~mT down to -65~mT and back. The behavior of
$R(H_a)$ is shown in Fig.~\ref{f2-RHRM}(c). From the positive field side at 15~mT, $R$
initially goes down, shows a small increase around 5~mT, but then goes down again,
followed by a steep dip in a very small field regime around $H_{dip}$ = -4.2~mT.
Reversing the sweep from the negative field side, the behavior is symmetric, with a
resistance dip now at +4.2~mT.
Also shown in Fig.~\ref{f2-RHRM} is the magnetization of the sample at 8~K, normalized by
the saturation magnetization $M_{s}$. The loop width is about 8~mT and the switch is
quite sharp, although some rounding can be seen close to the coercive field which is due
to the misalignment of the easy axis by a few degrees. It is clear that the deep dips in
$R(H)$ occur precisely at the switching field of the magnetic layer, which indicates that
the domain state of the ferromagnet is involved, which is most prominent in the steep
part of the magnetization reversal. Since the dip is very pronounced, we can also measure
$R(T,H_{dip})$ and compare this with $R(T,0)$ in Fig.~\ref{f1-RT}. We find that $R(T,
H_{dip})$ lies consistently {\it below} $R(T,0)$, with a maximum difference of about
10~mK. Several samples were measured which all show the same effect, and one data set for
a sample s/Py(20)/Nb(19) (reversed Nb and Py) is shown in Fig.~\ref{f3-P20N20}.
Again the deep dips in $R(H_a)$ (inset in Fig.~\ref{f3-P20N20}) are clearly present, but
there are differences in the details. For instance, $T_c$ is somewhat lower, around
4.85~K, presumably because Nb layer is slightly thinner. Also, the transition is sharper,
with a width of 50~mK. This necessitates a temperature stability during the field sweep
of better than 1~mK. The magnetization loop (not shown) is also wider, about 14~mT, and
the rise in $R(H_a)$ before the dip is reached is higher as well. We believe both facts
may be due to the fact that the rotation of the magnetization in the covered Py-layer
proceeds slightly differently than in the free but probably oxidized Py-layer. Again,
$R(T, H_{dip})$ lies consistently
{\it below} $R(T,0)$, with a smooth behavior of the difference. \\
The basic explanation we want to offer for our observations is that domain walls are
formed in the Py-layer during the switching of the magnetization, and that these domain
walls lead to the enhanced superconductivity. In order to argue this better, we performed
measurements on two other types of samples. Fig.~\ref{f4-alox}(a,b) show $R(H_a)$ data
(in the transition) on a {\it 'small'} (1.5~$\mu$m $\times$ 20~$\mu$m) sample
s/Nb(17)/Py(20) where no effect is found. The other is a 'large' sample
s/Nb(20)/AlOx(8)/Py(20), where a thin Al layer was deposited and oxidized before the Py
layer was grown. No dips are found, but instead a small resistance increase is seen in
the field region of the Py loop, probably due to the effect of stray fields from the
magnetic layer on the superconducting layer. This shows that a proximity coupling is
necessary for the dips to
occur. \\
Before discussing the effects of domain walls on the superconductivity in more detail, it
is necessary to address the characteristics of the walls. For the film thicknesses $d_F$
we use (around 20~nm), they are believed to be of N{\'e}el-type : the rotation of the
magnetization occurs in the plane of the sample rather than out-of-plane (Bloch-wall),
with the transition between the two at thicknesses around 40~nm
\cite{o'hand00,kneller62}. We note that the magnetic flux coming out of a Bloch wall
would  suppress the superconductivity rather than do the opposite. The other important
parameter is the width $\delta_w$ of the walls, which in Py is large due to the small
magnetic anisotropy. For Bloch walls, $\delta_w$ is of the order of 0.5~$\mu$m, for
N{\'e}el walls and the case that $d_F << \delta_w$ it is of similar magnitude. A detailed
study by scanning electron microscopy with polarization analysis on thick Py films with
surface N{\'e}el walls yielded 0.25~$\mu$m for the half-width of the walls
\cite{scheinf91}. The fact that $\delta_w$ is much larger than the low-temperature
coherence length $\xi_S$ of the superconductor (for our Nb, $\xi_S \approx$ 12~nm) will
come back in the discussion. The absence of an effect for the small samples
s/Nb(17)/Py(20) can be explained by the fact that no stable domains were formed during
switching of the magnetization direction. To confirm this assumption, we simulated the
switching behavior for a range of Py structures with thickness 20~nm and different size
and aspect ratio, using the OOMMFF code \cite{oommff}. Values for the saturation
magnetization $M_s$ and the magnetocrystalline anisotropy $K_1$ were determined from the
magnetization measurements and taken to be $M_s$~= 860$\times 10^3$ [A/m] and $K_1$~= 500
[J/m$^3$]. The easy axis direction was taken 6.5$^{\circ}$ away from the long axis of the
structure in order to take the slight sample misalignment into account . Some pertinent
results are shown in Fig.~\ref{fig5-domains}. In a structure with length $\ell$ and width
$w$ of both 10~$\mu$m, the magnetization loop is small and near the coercive field H$_c$
(around 2~mT) stable domain configurations are present. Upon decreasing $w$ and
increasing the aspect ratio $a = \ell / w$, stable configurations no longer occur above
$a$~= 2.5. For the 1~$\mu$m $\times$ 10~$\mu$m structure, without a stable domain
configuration, H$_c$ has increased to 10~mT. Note that this is still below the value
expected for a uniform rotation of the magnetization, which is of order $M_s/2$ according
to the Stoner-Wohlfarth model \cite{o'hand00}, since domains are formed during the
switching. The point is that they are not stable. Simulations on structures with larger
length (above 50 $\mu$m) showed stable multi-domain configurations near H$_c$ even for
structures with \textit{a} $>$ 10. It is therefore the combination of small dimensions
and large aspect ratio in our 'small' samples which precludes stable domains.

The mechanism we believe to be responsible for the enhancement of the superconductivity
in the 'large' samples is basically that the pair breaking experienced by a Cooper pair
is smaller when it samples different directions of the exchange field. In a sense, this
is the same mechanism as responsible for the F/S/F spin switch, but the simultaneous
sampling of two F-layers yields a different type of averaging. In order to estimate the
effect of the domain wall, we first consider the one-dimensional case of step-like
variation of $h_{ex}$ which induces changes in the order parameter over a distance
$\xi_F$ (the coherence length in the F-layer) leading to a lowered pair breaking
parameter in the vicinity of the domain wall \cite{buzdin84} and superconductivity which
is enhanced with respect to the depression of the uniform F-layer. The situation
resembles $T_c$-enhancement by twin planes \cite{khly87,abri88} and can be treated
accordingly. An estimate for the order of magnitude of the enhanced critical temperature
$T_{cw}$ can be made as follows. The variation in pair breaking occurring over a distance
$\xi_F$ induces a superconducting order parameter over a distance $\xi_S(T_{cw})$ around
the wall. In that case, the effective change in pair-breaking will be $\xi_F /
\xi_S(T_{cw})$, and the $T_c$-enhancement is correspondingly $\alpha \equiv$ $(T_{cw} -
T_{cF}) / T_{cF}$ $\approx \xi_F / \xi_S(T_{cw})$, with $T_{cF}$ the critical temperature
due to the homogeneous suppression of the F-layer. Taking into account that
$\xi_S(T_{cw})$ $\approx \xi_{S}(0) / \sqrt{\alpha}$, we obtain $\alpha \approx (\xi_F /
\xi_{S}(0))^2$. With $\xi_S(0) \approx$ 12~nm and $\xi_F$ for the strongly magnetic Py
$\approx$ 1~nm, the effect is in the range of 1~\%. The assumption of a step-like change
in $h_{ex}$ is not correct, given the large value for $\delta_w$, but it gives a feeling
for the orders of magnitude. For larger $\delta_w$, the effect will increase as long as
$\delta_w$ $< \xi_S(0)$, but for still larger $\delta_w$ it has to decrease to zero
again~: the Cooper pair cannot sense the variation in $h_{ex}$ any longer. In this limit,
the relative increase of $T_c$ is of order $(\xi_S(0) / \delta_w)^2$ \cite{buzdin03},
which is again about 1~\%. Another way of arguing that the experiment is sensitive to
domain wall formation comes from considering the temperature dependent (Ginzburg-Landau)
$\xi_S(T)$ directly. Since we are measuring very close to $T_c$, $\xi_S$ is actually much
larger than the low-temperature value. From a fluctuation analysis of $R(T)$ we estimate
$T_c$ to be close to the top of the transition. In Fig.~\ref{f3-P20N20} it would be at
4.83~K, where the 0-field and in-field curves start to separate. With a typical
transition width of 30~mK, the relative temperature at the zero of resistance $t_r = T /
T_c$ is $6 \times 10^{-3}$, which makes $\xi_S(t_r) = \xi_S(0) / \sqrt{1-t_r}$ about
0.15~$\mu$m. In the transition, therefore, the condition is $\xi_S(T) \approx \delta_w$,
and the Cooper pair samples a considerable part of the rotation of the magnetization.
This implies that in the case of Py the in-plane switch will only be visible close to
$T_c$, since at lower temperature the magnetization is homogeneous on the scale of
$\xi_S$. It is of interest to note that enhancement of critical currents below $T_c$ has
been reported for the case
of Nb/Co, where the domain walls are considerably smaller \cite{kinsey01}  \\

\noindent This work was supported by the "Stichting FOM". We thank A. Timofeev and S.
Habraken for assistance in some of the experiments.

\end{document}